\documentclass{article}
\usepackage[utf8]{inputenc}
\newcommand{\etal}{\textit{et al}. }
\newcommand{\ie}{\textit{i}.\textit{e}. }
\newcommand{\eg}{\textit{e}.\textit{g}. }
\usepackage{graphicx}
\title{Large-Scale Benchmarks for the Job Shop Scheduling Problem}
\author{Giacomo Da Col and Erich C. Teppan}
\date{January 2021}

\begin{document}
\maketitle

\section{Introduction}
The aim of this report is to present and provide access to two novel benchmarks for the Job Shop Scheduling Problem (JSSP). The JSSP is one of the most studied scheduling problem and as such, there are a large number of benchmarks available in literature (\eg \cite{lawrence,taillard,adams,storer,apple}). However, a common shortcoming of classic problem instances is the limited number of jobs and operations in comparison with real industrial scenarios. 

The industrial field is one of the domains that had more impact in the development of scheduling theory \cite{fisher,Blazewicz2007,baker2013}, to the point that even the terminology adopted by scholars derives from its semantic field. In fact, terms often used in the scheduling domain are strictly coupled with industrial concepts, like \emph{machines} to indicate resources and \emph{jobs} to indicate tasks. Following this terminology, the factory layout (\ie the number of machines and their functionality) is called \emph{shop}, and when a job is composed by sub-tasks ordered in a specific sequence, they are called \emph{operations}. Operations are linked to machines with the concept of \emph{operation type}. In fact, every operation has a type and each machine can process operations of certain types, and not others. 
 
Despite this strong link, scholars have begun to study the scheduling problem in a more abstract and ``pure'' form. This allowed researchers to concentrate on the aspects that are at the core of the problem complexity (\eg \cite{garey1976}), but made more and more difficult to apply the academic results on real-life scenarios \cite{fuchigami2017}. 

One of the aspects where this discrepancy is more visible is the size of the scheduling problems.In fact, one of the few benchmark targeted to industrial scenarios is the Taillard benchmark, from 1992 \cite{taillard}. He published a benchmark simulating the size of real industrial data, with the largest JSSP instances reaching 100 jobs to be scheduled on 20 machines. After his work, however, little effort was put into maintaining the parallelism between real industrial problems and JSSP instances. Nowadays it is not uncommon to reach scheduling problems as big as 1000 jobs on 1000 machines for a rather short planning horizon (e.g.\ a week); however, in JSSP there are no benchmarks of such size. Some recent studies, like Zhai \etal in 2014 \cite{zhai2014}, address the problem, but are still reluctant to test instances beyond 100 jobs on 50 machines. Others, like one proposed by IBM, mention tests done on large-scale scheduling instances that go beyond 1000 jobs, but said benchmark is not of public domain \cite{laborie2018ibm}. 

The experimentation of scenarios close to the real-world industrial applications is at the core of this report. Given the small size of the available benchmarks, we worked towards the definition of a new testing environment conforming to modern industrial standards. With this goal in mind, two novel large-scale benchmarks for JSSP were produced. 

\section{The Large-TA benchmark}
The first benchmark can be seen as an ``extension'' of the Taillard benchmark, meaning that the instances are created with the same procedure of \cite{taillard}, but in a larger scale. In fact, there are a total of 90 instances, divided in 9 groups of 10 instances each, spanning from $10 \times 10$ to $1000 \times 1000$, as shown in Table \ref{table:lt}. Following Taillard's specification, every job has to be processed by every machine and just once per machine. Thus, the number of operations of an instance corresponds to the number of jobs times the number of machines (instances of this type are typically referred to as ``rectangular''), peaking at one million operations to be scheduled, in the largest group. This benchmark was first introduced in a publication that compared two state-of-the-art constraint solvers on realistic industrial scheduling problems \cite{dacol2019cp}.

\begin{table}[h]
\centering
\begin{tabular}{rrr|c}
\textbf{Machs} & \textbf{Jobs} & \textbf{Operations} & \textbf{Instances} \\ \hline
10             & 10            & $100$                 & 10                 \\
100            & 10            & $1000$                & 10                 \\
1000           & 10            & $10\,000$                 & 10                 \\ \hline
10             & 100           & $1000$                & 10                 \\
100            & 100           & $10\,000$                  & 10                 \\
1000           & 100           & $100\,000$             & 10                 \\ \hline
10             & 1000          & $10\,000$               & 10                 \\
100            & 1000          & $100\,000$                & 10                 \\
1000           & 1000          & $1\,000\,000$             & 10                
\end{tabular}
\caption{List of the instances of the Large-TA benchmark, grouped by size. The last column indicates the number of instances for that group.}
\label{table:lt}
\end{table}

\section{The Known-Optima Benchmark} Contrary to the Large-TA, this benchmark does not comprehend rectangular instances, \ie a job can have less operations than the number of machines, and two operations of the same job are not forcibly processed by the same machine. 

\begin{figure}
\center
	\includegraphics[width=8.5cm]{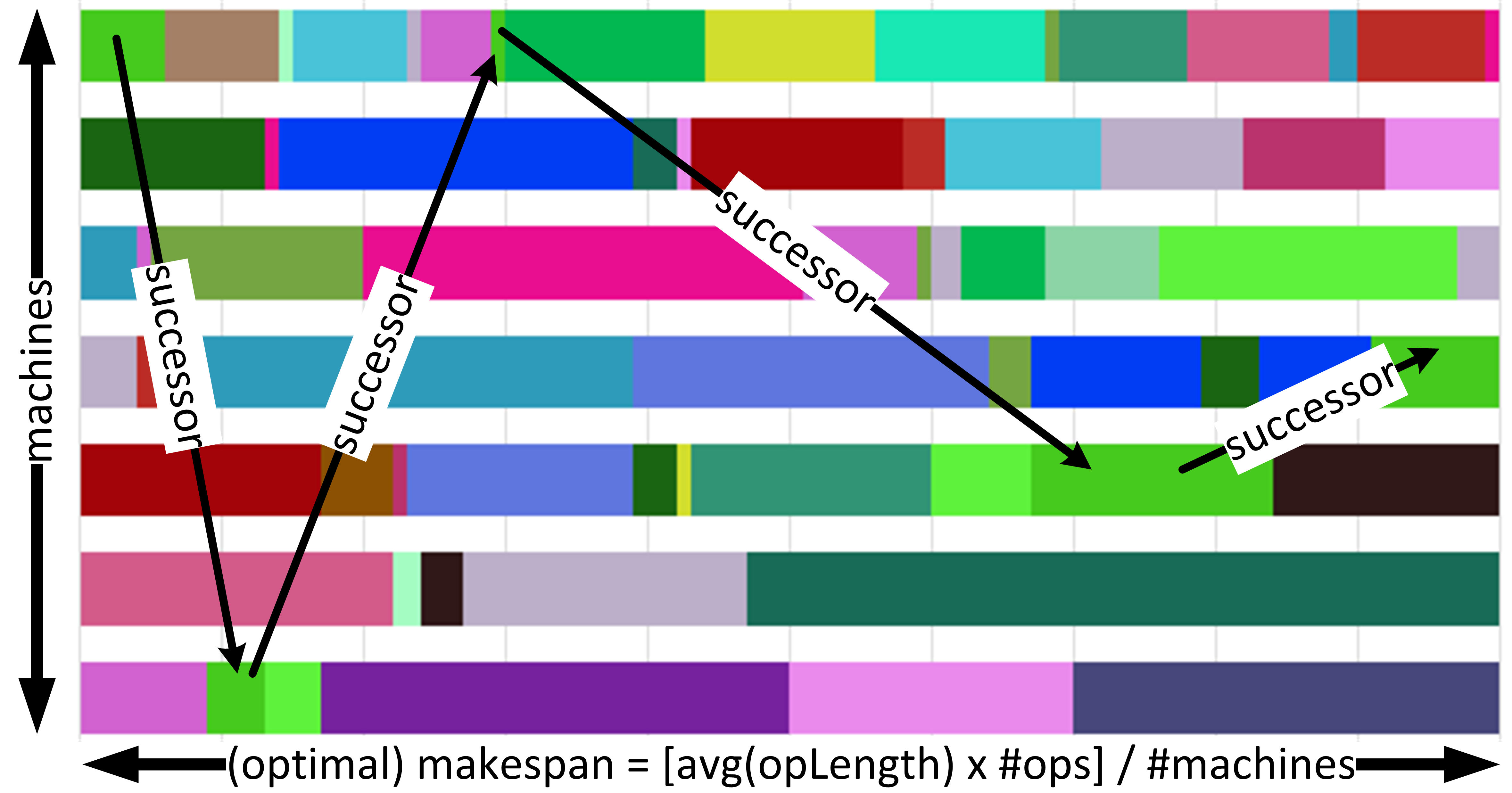}
	\caption{Principle of instance generation}
	\label{instanceGeneration}
\end{figure}

The creation process is as follows: First produce an optimal solution without (idle) holes for (1) a given number of job operations to be scheduled, (2) the number of machines and (3) the desired optimal makespan. This is done by randomly partitioning the machines' time continuum into the predefined number of partitions. Each partition corresponds to the processing period of one operation. Consequently, the optimal makespan, average operation length (avg(length)), number of operations ($\#ops$) and number of machines ($\#machines$) relate conforming to $makespan = \frac{\#ops \times avg(opLength)}{\#machines}$. Based on such a partitioning, successor relations are randomly generated. Each partition has maximally one successor and/or predecessor such that the successor's starting time is greater than the predecessors finishing time. Figure \ref{instanceGeneration} shows the principle. This creates scheduling problem instances where the best makespan is known before-hand, giving the opportunity to immediately check the quality of a given solution.

We applied two different procedures for generating random successor relations based on a pre-calculated solution:
\begin{enumerate}
\item \textbf{Short-jobs instances: }For each operation $op$ (in random order) define as successor $suc$ a random operation such that
\begin{itemize}
\item $suc$ is not on the same machine as $op$.
\item $suc$ starts later than $op$ ends.
\item $suc$ is not yet a successor of another operation.
\item If no such $suc$ exists $op$ has no successor.
\end{itemize}
\item \textbf{Long-jobs instances: }For each operation $op$ (in random order) define as successor $suc$ an operation such that
\begin{itemize}
\item $suc$ is not on the same machine as $op$.
\item $suc$ starts later than $op$ ends.
\item $suc$ is not yet a successor of another operation and
\item the time between $op$ ends and $suc$ starts is minimal.
\item In case that there are multiple possible successors, a random one is chosen.
\item If no such $suc$ exists $op$ has no successor.
\end{itemize}
\end{enumerate}

The two different generating approaches result in benchmark instances that are different in nature: (1) produces many jobs consisting of a small number of operations. In contrary, (2) produces fewer jobs but with a larger number of operations per job. This way, we can simulate complex products that need a lot of steps to be completed, as well as cases when there is a vast variety of simple jobs. The total amount of generated instances is 24 (12 short, 12 long). The total number of operations goes up to 100 thousand operations to be scheduled on up to one thousand machines. All instances have a minimal makespan of 600000, which is roughly a week in seconds, i.e.\ a very common planning horizon in the semi-conductor domain.



\begin{figure}[ht]
	\centering
		\includegraphics[width=1.1\textwidth]{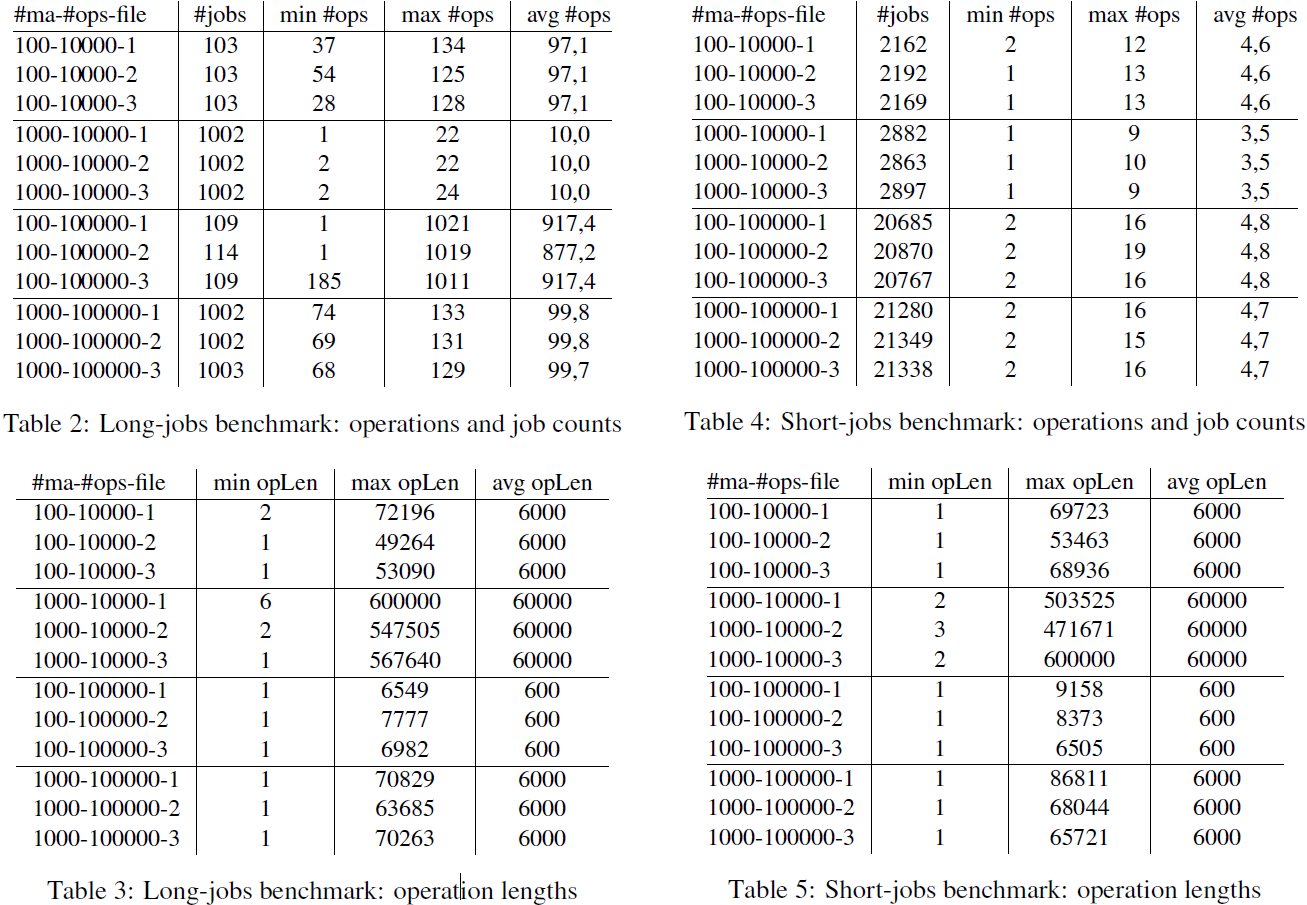}
	\caption{Collection of statistical elements about the two types of instances of the Known-Optima benchmark: long-jobs and short-jobs instances.}
	\label{fig:bench}
\end{figure}

The tables in Figure \ref{fig:bench} offer some statistical data on the instances. Tables 2 and 4 show the number of jobs and the min, max and average number of operation per job. The first group of three instances in Table 2 shows that in the long jobs, the max number of operations exceeds the number of machines. This means that there are jobs that are processed twice or more by the same machine. Similarly, there are jobs that go trough just a subset of the machines, since the min $\#ops$ is smaller than the number of machines. The second group has 1000 machines, and presents much shorter jobs on average. If we compare it with the same group of the short job instances in table 4, the long jobs have three times the operations on average (3.5 vs 10), compared to the 20 times multiplication factor of the first group (4.6 vs 97.1). The min $\#ops$ of instance 1000-10000-1 reveal that there are also jobs composed by a single operation even in the long-job instances.
Table 4 shows that in the short-job instances there are no cases where a job is processed by all machines, in fact, the max $\#ops$ is 19. The low avg $\#ops$ results in a high number of jobs (up to 20 thousand) with 3 to 5 operations on average. 

Table 3 and 5 show the distribution of the operation length of the operations. They vary a lot in both short and long job instances,between 2 to $600\, 000$. The optimal makespan on every instance is $600\, 000$, which represents roughly one week in seconds, to mirror real industrial scenarios, where it is typical to plan a whole week of schedule in one go. Given the optimal makespan at $600\, 000$, it means that in instance 1000-10000-1 long-jobs, there is at least one operation which is as wide as the whole optimal schedule. This benchmark is used is several recent works about industrial scheduling (\cite{teppan2018dispatching,dacol2019iclp,Teppan2020}).

\begin{figure}
	\centering
		\includegraphics[width=1\textwidth]{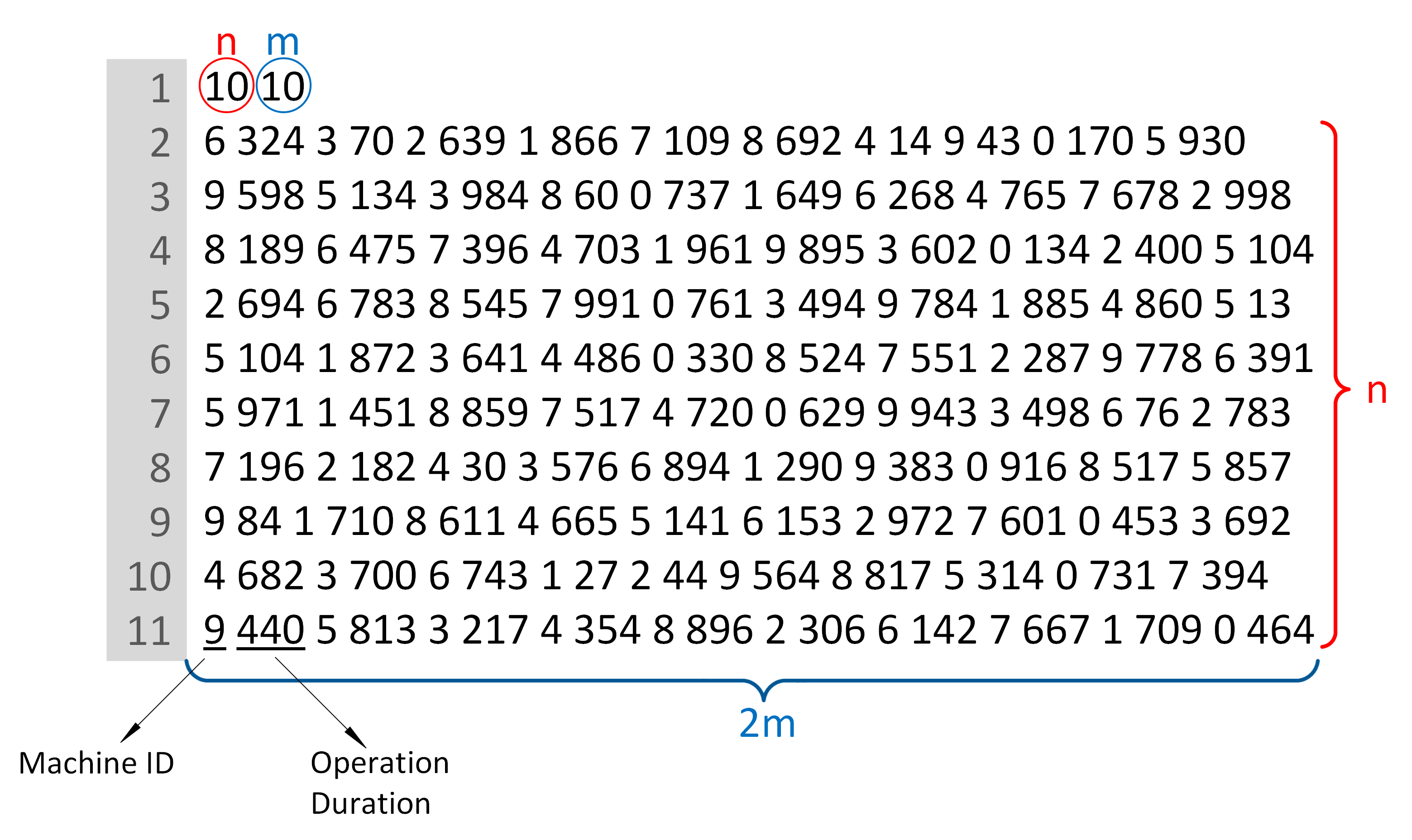}
	\caption{Explanation of the instance representation of the Large-TA benchmark. The instance shown is the file \emph{Large-TA/tai\_j10\_m10\_1.data}. Values in the grey field represent the line numbers.}
	\label{fig:instance}
\end{figure}

\section{File representation}
The instances are collected in two folders, one per benchmark. The representation of the JSSP instances is pretty similar in the two benchmarks. Figure \ref{fig:instance} shows the first instance of the Large-TA benchmark, with the description of the elements. The first line contains the size of the instance, with the number of jobs $n$ and machines $m$ respectively. Then, each line contains the description of one job and its $m$ operations. For each operation, it is specified the id of the machine where the operation is assigned to, as well as the duration of the processing time of the operation. 

\begin{figure}
	\centering
		\includegraphics[width=0.75\textwidth]{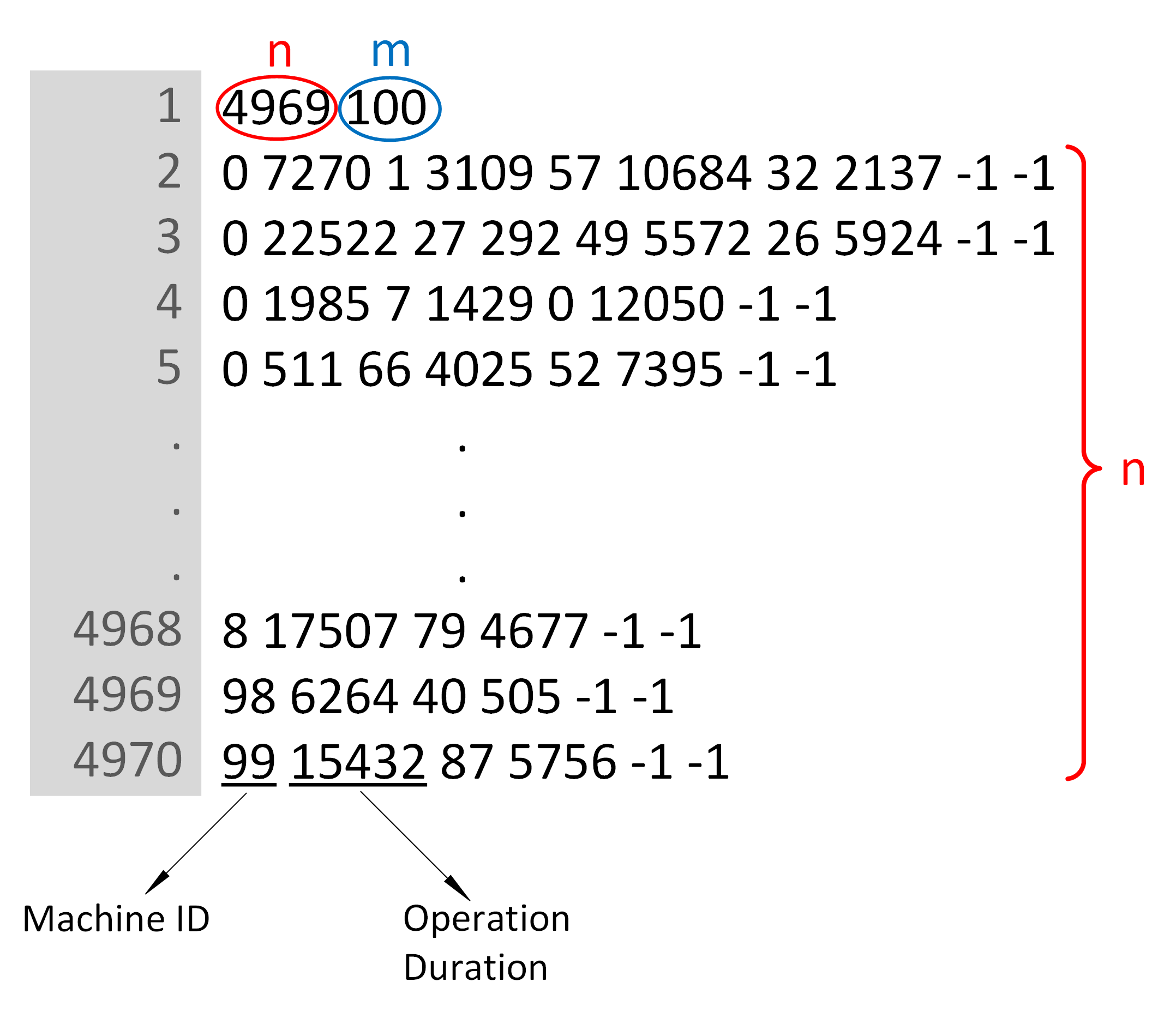}
	\caption{Explanation of the instance representation of the Known-Optima benchmark. The instance shown is the file \emph{Known-Optima/long-js-600000-100-10000-1.data}. Values in the grey field represent the line numbers.}
	\label{fig:instance2}
\end{figure}

Figure \ref{fig:instance2} shows the first of the Known-Optima benchmark. The definition of the jobs and operation is similar to the Large-TA instances. The only tangible difference is that, in this benchmark, the number of operations per job is not always the same, therefore, each line ends with the values -1 -1 to indicate the end of the operations' sequence. 

\begin{figure}
	\centering
		\includegraphics[width=0.6\textwidth]{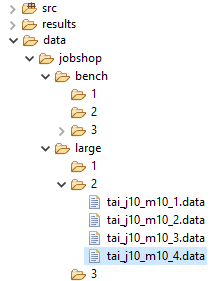}
	\caption{How the program expects the instances to be organized. In particular, ``bench'' refers to the Known-Optima benchmark, and ``large'' refers to the Large-TA benchmark. The benchmark is then divided in sub-folders, for easy split of the benchmark instances for parallel computation. There is no upper limit to the number of sub-folders, but there should be at least one. By convention, they are named after consequent numbers, but the name can actually be anything. Folder ``src'' contains the java source file, while ``results'' is the destination folder of the output of the computations.}
	\label{fig:folders}
\end{figure}

\section{Code for JSSP on OR-Tools and CP Optimizer}
In addition to the benchmarks, we provide some code snippets including the encodings of the JSSP on two Constraint Programming solvers (CP), namely CP Optimizer and OR-Tools. This code has been used to test the two solvers on the proposed benchmarks \cite{dacol2019cp,dacol2019iclp}. In the $cp\_solvers\_code$ folder we provide the source code in .java files. We \underline{do not} include the libraries of the solvers, because at least one of them is a proprietary solver that needs to be purchased\footnote{OR-Tools is available for free at https://developers.google.com/optimization. CP Optimizer can be obtained at https://www.ibm.com/analytics/cplex-cp-optimizer}. Figure \ref{fig:folders} explains the directory hierarchy to use for organizing the benchmark in input. Folder ``src'' contains the source files. For each solver, three encodings are available: naive, semi-naive and advanced. The difference between these encodings is explained in detail in \cite{dacol2019cp}. The java files are the following:
\begin{itemize}
\item Main.java: It's the launcher file;
\item MySolutionCallback.java: It's an extension of the SolutionCallback object, to gather the information needed when a solution is available;
\item SchedJobShop.java contains the code of the CP Optimizer solver for JSSP, advance encoding;
\item NaiveJobShop.java contains the naive version of the CP Optimizer solver encoding for JSSP;
\item SemiNaive.java contains the semi-naive version of the CP Optimizer solver encoding for JSSP;
\item SchedJobShopORTools.java contains the code of the OR-Tools solver for JSSP, advance encoding;
\item ORToolsNaive.java contains the naive version of the OR-Tools solver encoding for JSSP;
\item ORToolsSemiNaive.java java contains the semi-naive version of the OR-Tools solver encoding for JSSP;
\item SchedOpenShop.java contains a tentative encoding for the open shop problem (not tested);
\item INFO.java contains metadata information;
\item InstanceConverter.java: converts the instances from the MiniZinc format to the format described in section 4;
\end{itemize}

The code was tested on a Linux system with Ubuntu 18.04.3 LTS x86\_64. However, both solvers offer also Windows-compatible libraries. The libraries have to be referenced at command line when invoking java, using the commands \textbf{-cp} and \textbf{-Djava.library.path=}, or using the tools provided by IDEs. 

The java program takes 5 arguments:
\begin{enumerate}
    \item First argument (mandatory) selects the solver with a numeric value between 0 and 3:
    \begin{itemize}
        \item 0 - ORTools Advanced encoding
        \item 1 - CPOptimizer Advanced encoding
        \item 2 - ORTools Naive encoding
        \item 3 - CPOptimizer Naive encoding
    \end{itemize}
		\item Second argument (Mandatory) selects the dataset:
		\begin{itemize}
        \item 0 - Large-TA benchmark
        \item 1 - Known-Optima benchmark
        \item 2 - Classic instances benchmark, used in \cite{dacol2019cp,dacol2019iclp}
    \end{itemize}
		\item Third argument (Mandatory): Input sub-folder name
	    \item Fourth argument (Mandatory): Timeout of the solver
		\item Fifth argument (Mandatory): n of workers (threads) available to the solver
\end{enumerate}

\bibliographystyle{splncs03}
\bibliography{report}

\end{document}